\documentclass{aa}
\usepackage{graphicx}
\usepackage{natbib}
\bibpunct{(}{)}{;}{a}{}{,}
\newcommand{\Beins}{\mbox{1\hspace*{-0.085cm}l}}
\begin{document}

\title{Point field models for the galaxy point pattern}

\subtitle{Modelling the singularity of the two-point correlation
function}

\author{Martin\,Snethlage\inst{1} \and Vicent J.\,Mart\'\i nez\inst{2} \and
Dietrich\,Stoyan\inst{3} \and Enn\,Saar\inst{4}}

\offprints{Vicent J. Mart\'{\i}nez,
      email: Vicent.Martinez@uv.es}

\institute{Mathematical Department, Silesian University at Opava,
Bezru\v covo n\'am.~13, CZ-74601 Opava 1,
 Czech Republik
\and
Observatori Astr\`onomic, Universitat de Val\`encia,
E-46100 Burjassot, Spain
\and
Institut f\"ur Stochastik, TU Bergakademie Freiberg,
Bernhard-von-Cotta-Str.~2, D-09596 Freiberg, Germany
\and
Tartu Observatory, T\~oravere, 61602, Estonia}

\date{Received / Accepted}

\abstract{There is empirical evidence that the two-point
correlation function of the galaxy distribution follows, for small
scales, reasonably well a power-law expression $\xi(r)\propto
r^{-\gamma}$ with $\gamma$ between 1.5 and 1.9.
Nevertheless, most of the point field
models suggested in the literature do not have this property.\\
This paper presents a new class of models, which is produced by
modifying point fields commonly used in cosmology to mimic the
galaxy distribution, but where $\gamma=2$ is too large. The points
are independently and randomly shifted, leading to the desired
reduction of the value of $\gamma$.
        \keywords{galaxies: statistics --- large-scale structure of
        universe --- methods: statistical}}

\maketitle

\section{Introduction}
The two-point correlation function $\xi(r)$ has been the primary
tool for quantifying large-scale cosmic structure
\citep[see][]{Peebles80}. Several estimates of this statistical
quantity from redshift surveys suggest a power-law shape of
$\xi(r)$ for $0.5\,h^{-1}\mathrm{Mpc}<r<10\,h^{-1}\mathrm{Mpc}$
($h$ being the Hubble constant in units of 100 km s${}^{-1}$
Mpc${}^{-1}$)
\begin{equation}\label{powerlaw}
        \xi(r)=\left(\frac{r}{r_0}\right)^{-\gamma},
\end{equation}
with $\gamma$ between $1.5$ and $1.9$, see for example
\citet{Davis83}, \citet{Maddox90}, \citet{mart93} or
\citet{Pons99}. Thus, $\xi(r)$ seems to have a singularity of
order $\gamma$. However, taking into account that galaxies
have a physically finite extended size, the shape of $\xi(r)$ has
necessarily to deviate, at very small scales, from the power-law
behavior.

It is important to be able to generate controllable point
processes resembling the galaxy distribution, not only showing
visual similar patterns to the observed one, but also reproducing
the same statistical properties. These processes are necessary to
test the performance and, in particular, the discriminative power
of the statistical estimators used to describe the cosmic texture
\citep{marsaa02}. It is also worth mentioning that two point
fields could present similar second-order characteristics, such as
the two-point correlation function analyzed in this paper, but
show completely different results for higher order statistics.

Several methods have been published so far to generate mock
catalogues to mimic the galaxy distribution. We could distinguish
two main kinds of methods: Those based on a physically motivated
procedure and those based on a stochastic model with controllable
statistical properties. The first family of methods 
includes the following two main approaches:
\begin{enumerate}
\item
Starting with cosmological $N$-body simulations based on the gas--dark
matter dynamics, one can extract galaxy catalogues by applying
physical schemes related with the galaxy formation processes 
\citep{kau99}. This is obviously the
most realistic way to obtain mock representations of the galaxy
distribution. However, it requires much computing time and
therefore the cost of obtaining several realizations is high.
\item
The second possibility is to treat the dynamics of the non-linear
clustering of matter by means of approximate methods based on
higher order perturbation theory. A recent example is
the so-called PThalo model introduced by \citet{sco02}. These
authors select virialized dark matter halos from 
simulations obtained by
applying the second-order Lagrangian perturbation theory. Then, given
a biasing scheme which accounts for the fact that the galaxy
formation needs not to be proportional to the local matter
density, one can extract realistic galaxy distributions.
\end{enumerate}

The second family of models for the galaxy distribution are those
which are based on statistical or geometrical prescriptions. These
models are more flexible than those from the first family, can be 
easily simulated and manipulated, and have
more controllable statistical properties. In many cases, these
statistics can be calculated analytically. 
The first model of this family was introduced by
\citet{neyman} and became very popular in spatial statistics.
The characteristics of this model were probably prefigured by the
Lick survey. \citet{Peebles80} reviews two point processes
providing clustering patterns with power-law behavior for the
two-point correlation function: The Soneira and Peebles
hierarchical model \citep{sp78} and the Mandelbrot's
prescription based on the Rayleigh-L\'evy random walk
\citep{man82}. 
The first model, based on a superposition of
several fractal clumps, reproduces well the second-, third-, and
fourth-order correlation functions. In the Rayleigh-L\'evy
model, galaxies are placed at the steps of a random walk starting from
an initial position, where each jump takes an isotropically chosen
random direction with the length following a power-law probability
distribution function. Although this model shows point
distributions with little resemblance to the observed one, it has
been successfully applied to test the error measurements of
counts-in-cells statistics \citep{sza96}.
The singularity of the correlation function in these models
can be varied in a rather wide interval.

Other point field models published in literature present a
two-point correlation function without any singularity or with a
singularity of order $2$. A well-known example is the Cox segment
process, where points are situated on randomly oriented line
segments. Its correlation function was derived by \citet{Stoyan95}
and has a singularity of order 2. A similar cluster process was
introduced by \citet{Buryak96}, where the points (i.\,e., the
galaxies) populate a mixture of randomly oriented lines and
planes. The singularity of the correlation function is determined
by the line component and has again the order 2.

\citet{Weygaert91} suggests
a point field model which is built from the vertices of a Voronoi
tessellation calculated from a randomly distributed nuclei. This
model, based on one of the most interesting constructions studied
in stochastic geometry, has the property of mimicking the
statistical properties of the distribution of clusters of
galaxies. However, for this model, the order of the singularity of $\xi(r)$
is again 2.

A still simpler point field model is appearing in
\citet{Stoyan92}. It is a Poisson cluster process where each
cluster consists of only two points. The corresponding two-point
correlation function can be given a singularity of any desired
order between 0 and 3. Of course, this point field is rather
primitive and has no physical motivation. It makes sense only with
regard to the singularity of the two-point correlation function.

This paper investigates an idea which leads to a new class of
point field models. They can have any desired
value of $\gamma$ between 0 and 3 and seem to be more realistic than the previous
models.

The models listed above are based on simplifying
assumptions which are not well fulfilled in reality. The idea is now to
model the difference between ideal and real world by means of an
additional random element, i.\,e., by giving independently a
random shift to each point of the theoretical model.

The analytical investigation of the effect of these random shifts
is very difficult and is therefore carried out only for the case
of relatively simple Poisson cluster processes, the Neyman--Scott
processes \citep[see][for a description of these
processes]{marsaa02}. It will be shown that a suitable shift of the points
will reduce $\gamma$. However, the shift may also be too big and
completely destroy the singularity of $\xi(r)$. This is the case,
for example, if the shift is Gaussian distributed.

In case of the Voronoi vertices process only simulation based
investigations are possible. A segment Cox process is also studied
by this means. The analysis performed on both point processes
suggest that the effect on $\gamma$ is similar to that for
Neyman--Scott processes.

\section{Point field models}\label{ppdne}
This section presents some point field models whose pair
correlation functions have a singularity at $r=0$. All of them
have considerably deficiencies with regard either to the exponent in
the power-law (\ref{powerlaw}) or to the
underlying physics.

\citet{Buryak96} suggested a cluster process, where the galaxies
are situated on randomly oriented lines and planes of infinite extent.
It models the observed
fact that galaxies often seem to be situated on filamentary
and planar structures \citep{mst}. The cluster points form one-dimensional
Poisson processes on the lines and two-dimensional
Poisson processes on the planes. The two-point correlation function
of this point field is proportional to $r^{-2}$ for
small $r$, thus $\gamma=2$.

A similar process has been introduced by \citet{Pons99} for
testing the performance of different estimators of the two-point
correlation function. A segment Cox point field was used for
this purpose. The process is generated by first placing randomly, with
a given density $\lambda_s$,
segments of fixed length $\ell$, within the window. Secondly,
points are scattered randomly on the segments with a given value
of the mean number of points per unit length on a segment, $\lambda_\ell$.
\citet{Stoyan95} have calculated the two-point correlation function
of this model which reads

\begin{equation}
\xi(r) = {1 \over 2 \pi r^2 \lambda_s \ell}-{1 \over 2 \pi r
\lambda_s \ell^2}
\end{equation}

Again, a value of $\gamma=2$ describes the point process built by
the vertices of a Poisson--Voronoi tessellation, see \citet{Icke87}
and \citet{Weygaert89}. The assumption that $\gamma$ is equal to
2 has been verified by \citet{Heinrich98}.

In all these models $\gamma$ is too large.  A rather simple point field
model which may have any $\gamma$ between 0 and 3 is presented by
\citet{Stoyan94}. The points of a homogeneous Poisson process of
intensity $\lambda_p$ define the centres of independent clusters.
Each cluster consists of two points. The two-point correlation
function is
\begin{equation}
        \xi(r)=\frac{f(r)}{8\pi\lambda_pr^2},
\end{equation}
where $f(r)$ is the probability density function of the distance
between the two cluster points. This point field is a member of
the class of Neyman--Scott processes, see Sect.~\ref{poivvcv}. The
size of $\gamma$ depends on the asymptotic behaviour of $f(r)$
for $r\to0$. If $f(r)$ behaves as $r^{\alpha}$, $\gamma$ is
equal to $2-\alpha$. Therefore, a suitable choice of $f(r)$ leads
to any desired order. Unfortunately, this point field is very
primitive and has no physical motivation. Thus, it is unsuitable
to model the galaxy point pattern and is only of technical value.
An interesting generalization of this model --the so-called
Gauss--Poisson process-- has been presented and analyzed
by \citet{Kerscher01}. \citet{Kersetal} have shown that this
model cannot describe well the distribution of the galaxy clusters.

All of the point fields investigated until now have considerable
deficiencies. Those which have the right $\gamma$ are quite
unrealistic models for the galaxy point pattern. And those which
are appropriate in some astronomical sense have the wrong
$\gamma$. The next section provides a method which yields
physically motivated models with a two-point correlation function
which behaves as desired.

\section{The effect of independent random shifts}\label{hasekase}

\subsection{Arbitrary shifts applied to Neyman--Scott processes}
\label{poivvcv}
This subsection deals with a class of relatively
simple Poisson cluster processes, the so-called Neyman--Scott
processes \citet{neyman}. They result from homogeneous clustering
applied to a Poisson process. The points of a homogeneous Poisson
process of intensity $\lambda_p$, called parent points,
define the centres of clusters.
The clusters are independently and
identically distributed. A cluster consists of a random number of
independently and identically distributed daughter points. The
random distance between a daughter point and its corresponding
parent point is distributed according to a probability density
function $d(r)$. The two-point correlation function of such a
process is proportional to $f(r)/r^2$, see, for example,
\citet{Stoyan95}. Here, $f(r)$ is the probability density function
of the distance between two randomly selected daughter points of
the same cluster.

Given such a Neyman--Scott process, we now apply
independent isotropic random shifts to all its points.
The isotropic random shift
consists of shifting a point by a random length in a uniformly
distributed direction. The random length of the shift
has a probability density function $d^*(r)$. For example,
in case of Gaussian shifts $d^*(r)$ is the Weibull density function
with $p=2$ and $b=1/(2\sigma^2)$
\begin{equation}
        \frac{r^2}{\sigma^2}\exp \left( -\frac{r^2}{2\sigma^2} \right ),\quad
        r\geq0
\end{equation}
(this is the probability density function of the length of a
Gaussian distributed vector in 3-D).

The random shift changes the distribution of the interpoint
distance between two points of the same cluster. We denote
the corresponding distribution function by $F^*(r)$,
and the probability
density function by $f^*(r)$. The resulting two-point correlation
function is proportional to $f^*(r)/r^2$.

The aim of the following is to determine the asymptotic behaviour
of $f^*(r)$ for $r\to0$. This will yield a recipe how to choose
the function $d^*(r)$ in order to obtain any desired $\gamma$.

\begin{figure}
 \centering
 \includegraphics[width=6cm]{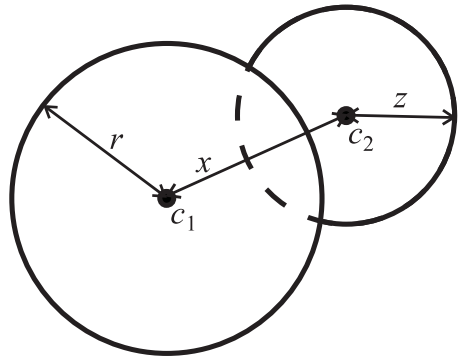}
 \caption{The quantity $F^*(r)$, demonstrated for the
two-dimensional case. The initial distance between the two cluster
points $c_1$ and $c_2$ is $x$, the length of the shift is $z$. The
distance between the shifted and unshifted points is smaller than
or equal to $r$ when the shifted point belongs to the dashed
line.}
 \label{hase}
\end{figure}

For a first step, assume that every cluster contains exactly two
points and that only one of them is shifted. $F^*(r)$ is then the
probability that the distance between both points is smaller than
or equal to $r$. For example, let the initial distance between both
points be $x$ and let the shift of length be $z$, see Fig.~\ref{hase}.
Then, $F^*(r)$ is equal to the fraction of that part of a sphere
of radius $z$ centred on the shifted point, which is not farther
from the unshifted point than $r$.

As the initial distance between the two cluster points is
distributed according to the probability density function $f(r)$
and as the length of the shift is distributed by $d^*(r)$,
we can write
\begin{equation}\label{inti}
        F^*(r)=\int\limits_{-\infty}^\infty\int\limits_{-\infty}^\infty
 f(x)d^*(z)\frac{\mathrm{A}(x;z;r)}{4\pi z^2}\textrm{d}x\textrm{d}z,
\end{equation}
where $\mathrm{A}(x;z;r)$ is the two dimensional volume of the
intersection of a ball of radius $r$ with a sphere of radius $z$,
whose centres are at the distance $x$:
\begin{eqnarray}
        \lefteqn{\textrm{A}(x;z;r)} \nonumber \\
 & =\left\{
        \begin{array}{ll}
                4\pi z^2, & \mathrm{if} \enspace x<r-z \enspace\mathrm{and}
                \enspace r\geq z\\
                \frac{\displaystyle\pi z}{\displaystyle x}
                \left(r^2-(x-z)^2\right)& \mathrm{if} \enspace
                \vert r-z\vert\leq x\leq r+z\\
                0 &  \mathrm{otherwise}
        \end{array}\right.
\end{eqnarray}
(the dashed line in Fig.~\ref{hase}).

It is rather complicated to find the integral (\ref{inti}); we
describe it in the
Appendix. The result is that the behaviour of $f^*(r)$ for small
$r$ depends only on the corresponding behaviours of $d^*(r)$ and
$f(r)$. If $d^*(r)$ behaves as $r^\alpha$ and $f(r)$ as
$r^{2-\gamma}$ (where $\gamma$ is the order of the singularity
of the correlation function before the shifts), then $f^*(r)$ behaves as
$r^{3-\gamma+\alpha}$, i.\,e., the order of the singularity of
$\xi(r)$ after the shifts is $\gamma-(\alpha+1)$. It has been
reduced, as $\alpha+1>0$ (else, $d^*(r)$ cannot be a
density function).

Now, let the number of daughter points per cluster be arbitrary
and let all of them be randomly shifted. As the daughter points
are independently distributed, this situation is principally equal
to that one before, only with an initial $f(r)$ which behaves as
$r^{3-\gamma+\alpha}$ instead of $r^{2-\gamma}$ (that is the
resulting density function after randomly shifting the first
points). It follows immediately that $f^*(r)$ behaves as
$r^{3-\gamma+\alpha+(\alpha+1)}=r^{4-\gamma+2\alpha}$. This
corresponds to a singularity of $\xi(r)$ of order of
$\gamma-2(\alpha+1)$.

Thus, power-law distributed random shifts with
density $d^*(r)\sim r^\alpha$ ($-1<\alpha<0$) will reduce the
order of the pole of the correlation function
by $2(\alpha+1)$.

\subsection{Shifting simulated point processes}
\label{iieeo}

The Neyman-Scott processes are too simple to
describe the observed galaxy distribution.
As we have not found analytic expression for the probability
density functions for more realistic point processes,
we have studied the effect of random shifts
on such processes by means of simulations.

\begin{figure}
 \centering
\resizebox{.35\textwidth}{!}{\includegraphics*{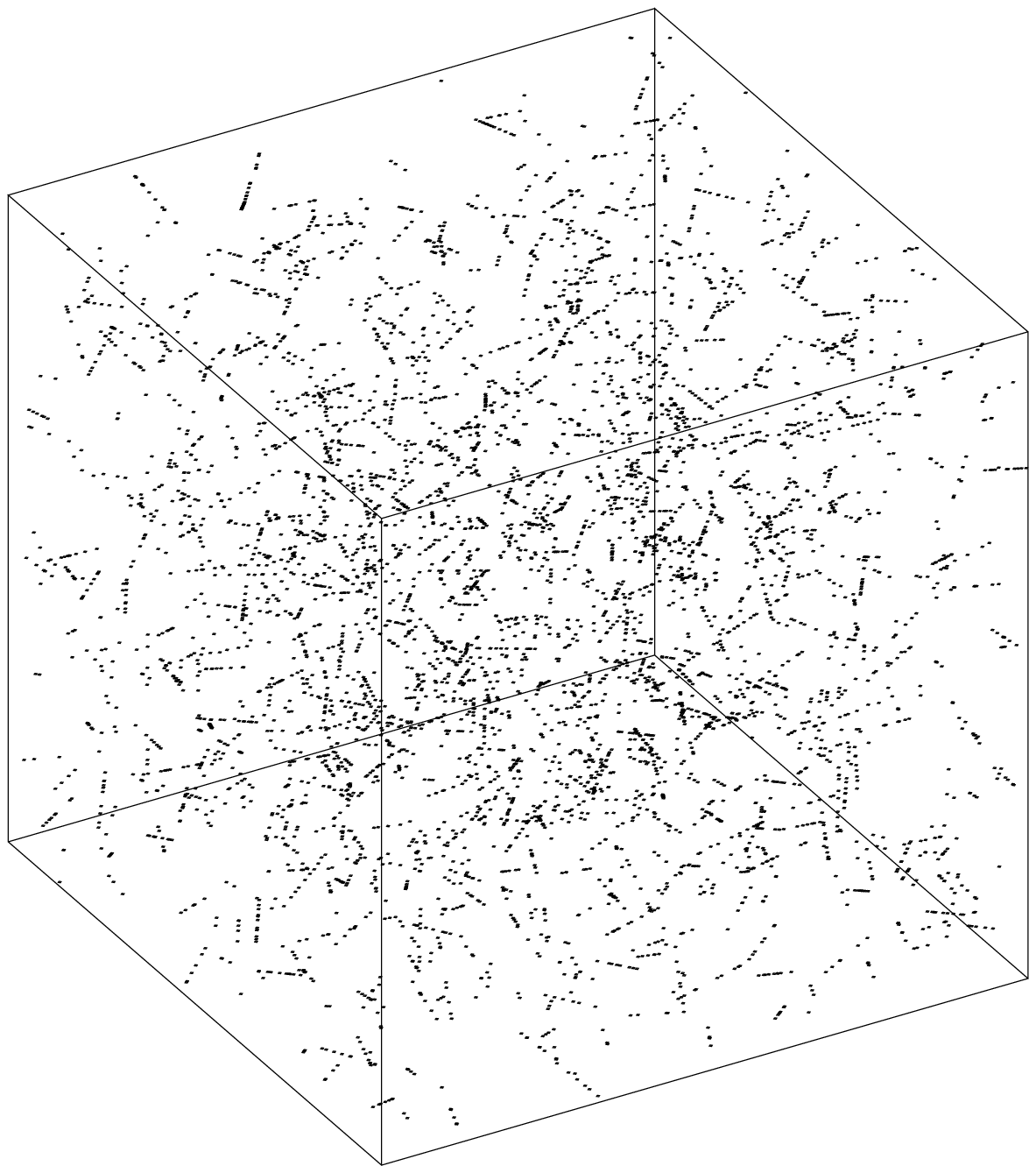}}\\
\resizebox{.49\textwidth}{!}{\includegraphics*{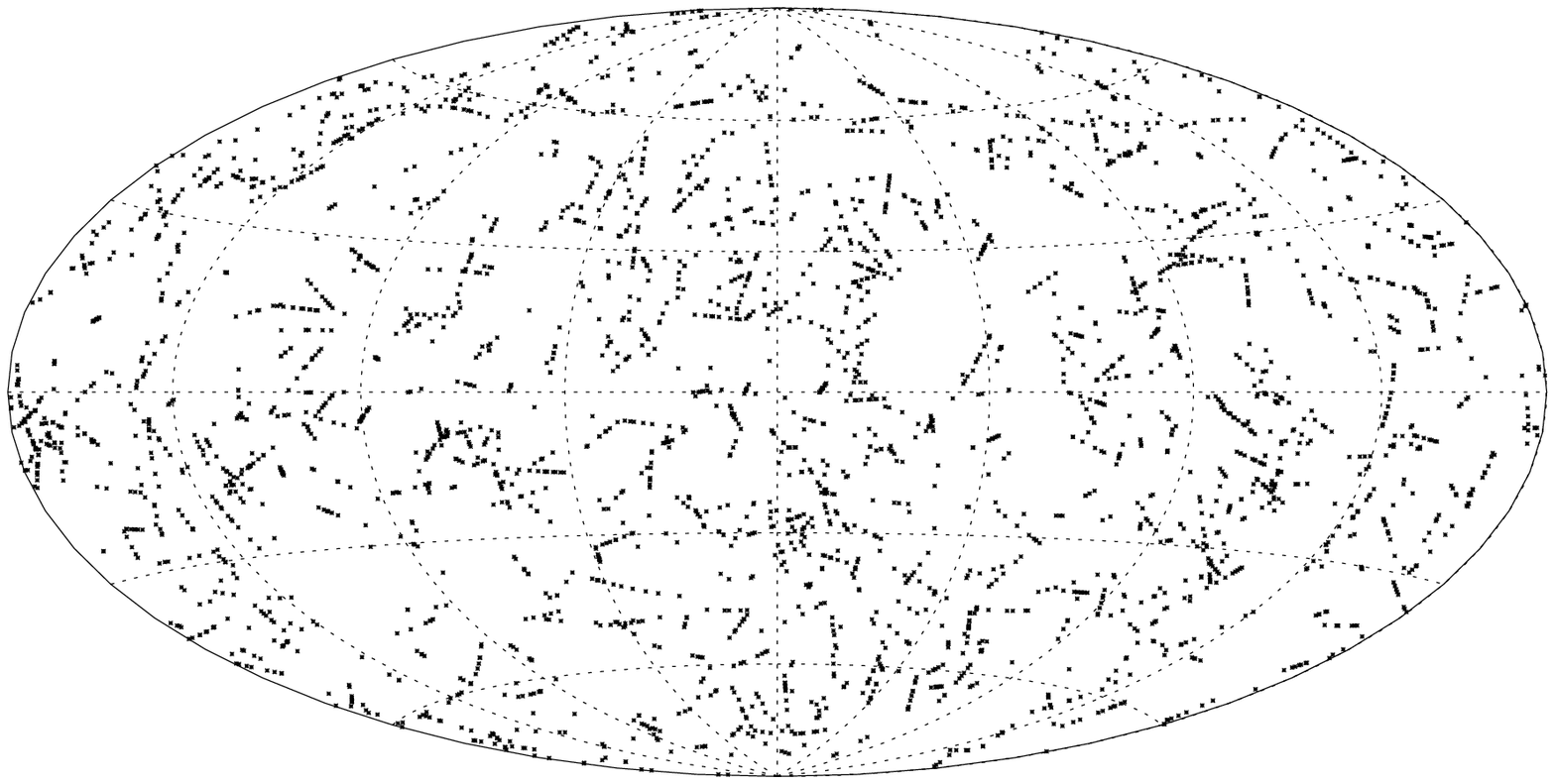}}
 \caption{The upper panels show a realization of the
segment Cox process. The bottom panels show
the Hammer--Aitoff projection of the points lying within
the maximal sphere included in the cube.}
\label{cubescox}
\end{figure}

        The first
model to analyze is the Cox segment process introduced in
Sect.~\ref{ppdne}. The realization of the model has been
simulated in a box of side $L=100$ with parameters
$\lambda_s=10^{-3}$, $\lambda_\ell=0.6$, and $\ell=10$
(see Fig.~\ref{cubescox}).

\begin{figure}
 \centering
\resizebox{.35\textwidth}{!}{\includegraphics*{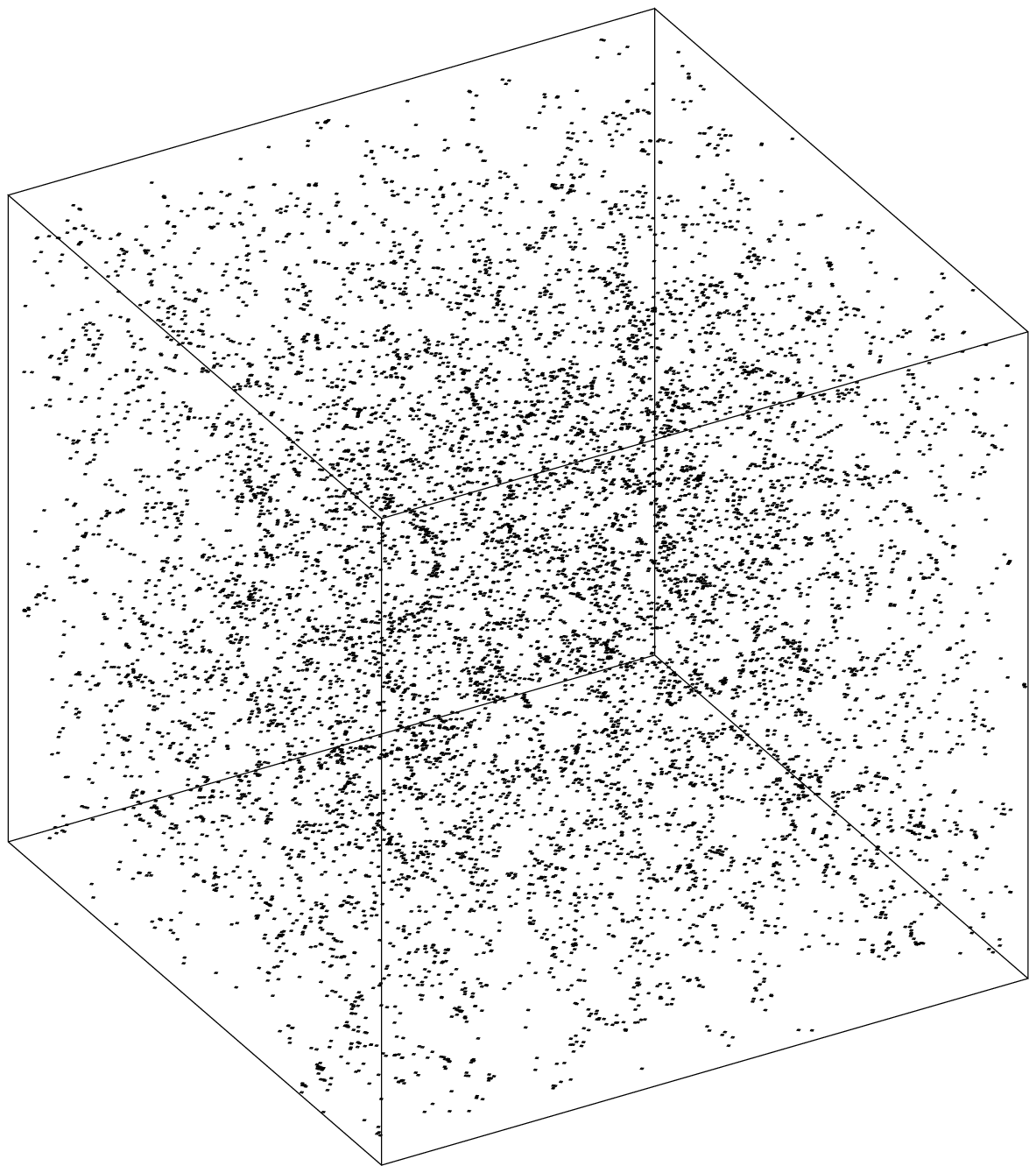}}\\
\resizebox{.49\textwidth}{!}{\includegraphics*{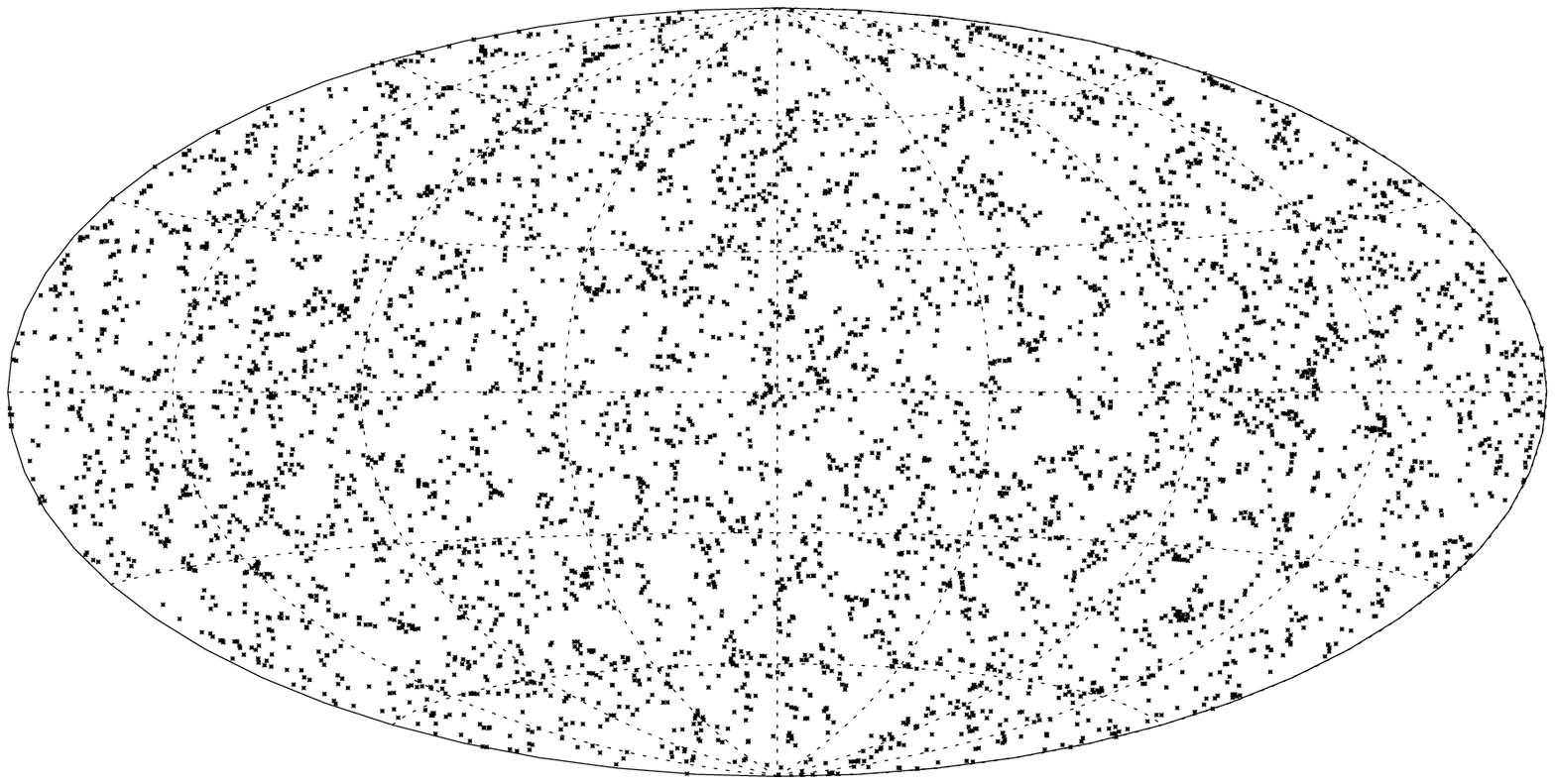}}
 \caption{The upper panels show a realization of the
a Poisson--Voronoi vertices process. The bottom panels show
the Hammer--Aitoff projection of the points lying within
the maximal sphere included in the cube.}
\label{cubesvor}
\end{figure}

The second model is based on Voronoi tessellations.
\citet{Weygaert91,Weygaert89,Icke87} suggested that clusters of
galaxies, the peaks of the matter density field, are  situated at
the vertices of the tessellation. It is remarkable how well this
model describes the cluster distribution. When nothing else is
known, one can assume a completely random distribution of the
initial points, the nuclei that will become the centers of the
Voronoi cells, i.\,e., the empty regions. Since the initial points
form a homogeneous Poisson process, the tessellation is usually
referred as a Poisson--Voronoi tessellation. Fig.~\ref{cubesvor}
shows a realization of this process consisting of 10085 vertices
resulting from the Voronoi tessellation of 1500 Poissonian
distributed nuclei within a box of side 100.

The determination of the two-point correlation function of the
corresponding vertices process is very complicated.
\citet{Heinrich98} have shown that the singularity of $\xi(r)$ at
$r=0$ is of order 2. As studies of the
real galaxy point distribution suggest an order between 1.5 and
1.9, the vertices process cannot strictly be an appropriate model
for the galaxy distribution.

A quite natural way to compensate for this difference between the
model and reality is to add perturbations to each point of the
Voronoi galaxy model, as we did for the Neyman-Scott models
in the previous section. That means that, after realizing a sample of
this model, each point is independently and randomly shifted. This
has the additional effect that the close attraction between the
points is relaxed. As it seems to be plausible that the degree of
the attraction between the points corresponds to the order of the
singularity, the relaxation should lead to a reduction of this
order.

Certainly, there is no prior theoretical distribution for the
random shifts. But the results of the previous section suggest
that a power-law distribution of shifts should work well.
There we determined the effect of arbitrary random
shifts on the order of the singularity of $\xi(r)$ in case of
Neyman--Scott processes. As these processes are based on a
homogeneous Poisson process as well as both the Cox segment
processes and the vertices process of a
Poisson--Voronoi tessellation, it is plausible to assume that in all
cases the effect of random shifts is similar. Thus, also in case of the
simulated processes, any suitable
density function $d^*(r)$ should
lead to a reduction of the order of the singularity.

\begin{figure}
 \centering
\resizebox{.24\textwidth}{!}{\includegraphics*{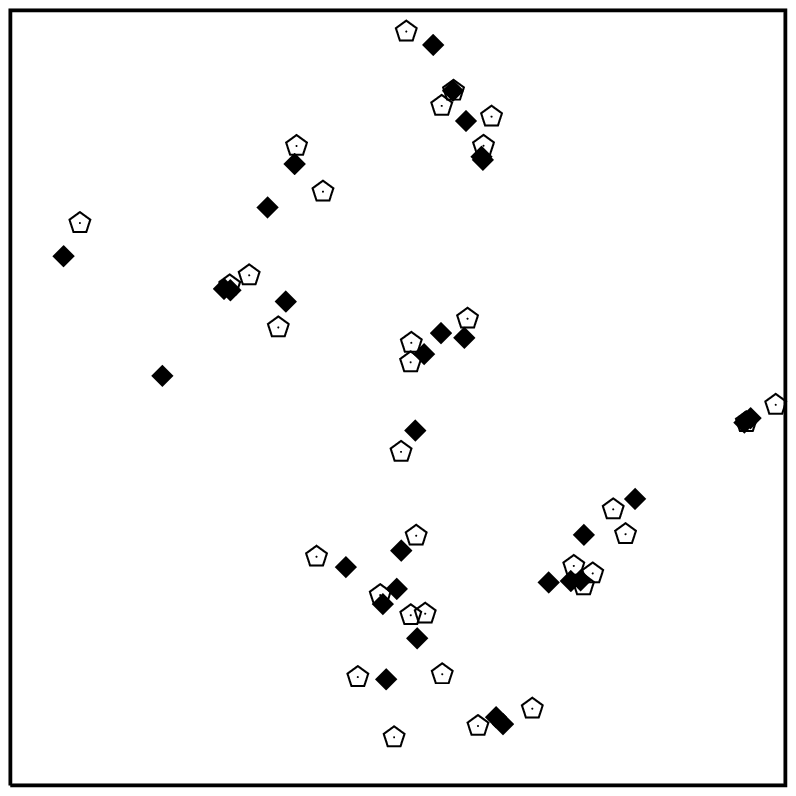}}
\resizebox{.24\textwidth}{!}{\includegraphics*{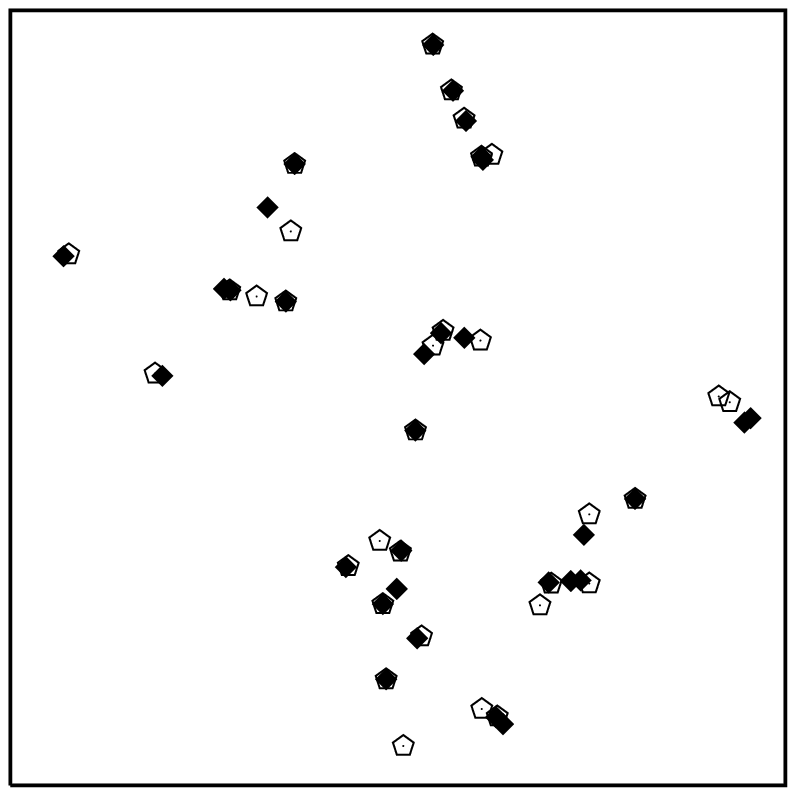}}\\
\resizebox{.24\textwidth}{!}{\includegraphics*{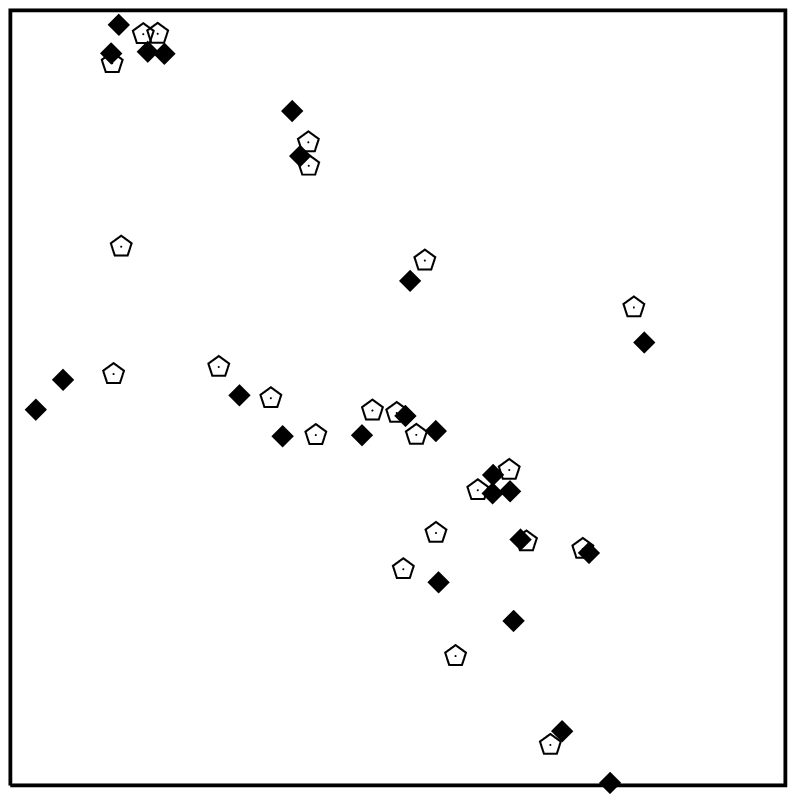}}
\resizebox{.24\textwidth}{!}{\includegraphics*{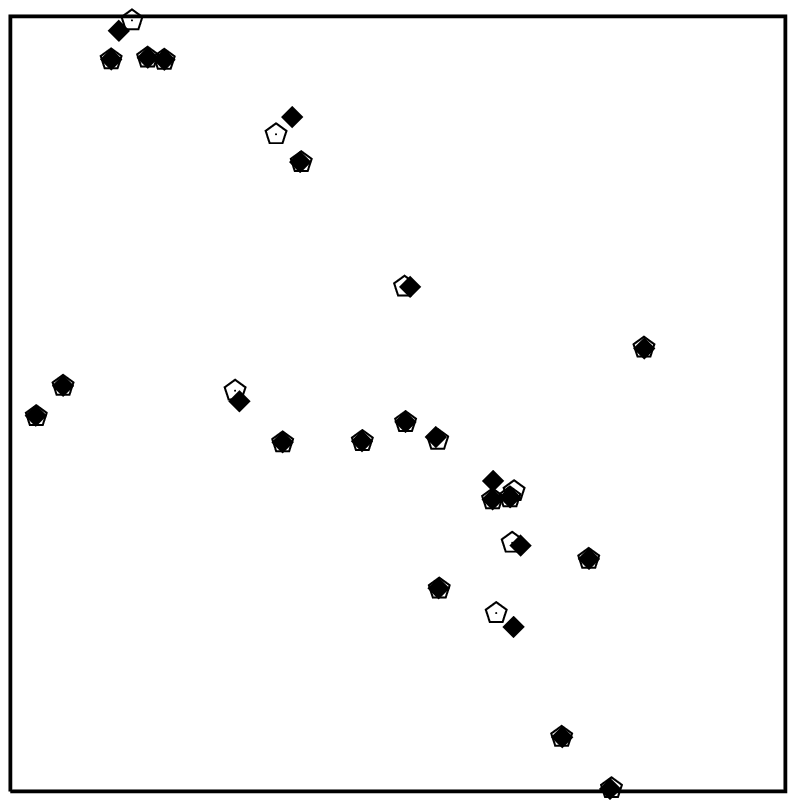}}\\
 \caption{Shifting Cox and Voronoi processes. The upper panels
show the orthogonal projection on the bottom of a small parallelepiped,
with dimensions $15\times15\times20$,
drawn from the Cox segment process (Fig.~\ref{cubescox}),
the lower panels show the projection on the bottom of
a parallelepiped with dimensions
$15\times15\times10$, drawn
from the Voronoi vertices
process (Fig.~\ref{cubesvor}). The left panels demonstrate Gaussian shifting,
the right panels demonstrate power-law shifting. The
solid symbols represent the original process, the open symbols
show the shifted points.}
 \label{points}
\end{figure}

We study this conjecture by simulation. We started with
the simulated Cox segment process and the Poisson-Voronoi
vertices process and applied random shifts.
The random shifts used
are distributed according to the probability density function
\begin{equation}
d^*(r)=\left\{
        \begin{array}{l@{\quad}l}
        0.25r^{-0.75} & \mbox{for }0\leq r\leq1,\\
        0 & \mbox{otherwise}.
        \end{array}
        \right.
\end{equation}
The effects of the power-law shifts are demonstrated in Fig.~\ref{points}
(right panels).

For comparison, we also tried shifting the points by
Gaussian-distribution shifts. As Gaussian distribution
is commonly used
when modelling an unknown random effect, it can serve as
a useful reference model.
Thus, given the simulated point fields, both the Cox segment process and
the Poisson--Voronoi vertices process,
every point is independently shifted by a
three-dimensional Gaussian distributed vector with $\sigma=0.5$.
Fig.~\ref{points}
(left panels) shows the result of this procedure.

Correlation functions amplitudes usually demand averaging
over a rather large number of pair distances and are not
easy to use near the pole. The expected power-law behaviour
can be analyzed better using
the correlation integral -- the so-called
$K$-function,
\begin{equation}
K(r) = \int_0^r 4 \pi s^2 [1+\xi(s)]\textrm{d}s.
\end{equation}

If $K(r)$ behaves as a power law, we can define
the correlation dimension $D_2$ as the exponent of the relation
$K(r) \propto r^{D_2}$.

We used Ripley's estimator for a cuboidal window \citep{Baddeley93}:
\begin{equation}
\hat{K}(r)=\frac{V}{N^2}\sum_{i=1}^N\sum_{j=1 \atop\scriptstyle j\ne i}^N
        \frac{\theta(r-|\mathbf{x}_i-\mathbf{x}_j|)}{\omega_{ij}},
\label{Ripley}
\end{equation}
where $V$ is the volume of the cube, $N$ is the total number
of points, $\theta(\cdot)$ is the Heaviside's step function
and the weights $\omega_{ij}$ account for the edge correction,
being the conditional probability that point $j$ would be observed
given that it lies at distance $r$ from point $i$.

If we find all the distances $r_{ij}=|\mathbf{x}_i-\mathbf{x}_j|$
and the corresponding weights $\omega_{ij}$, we can
order them by increasing distance, and define the
ordered set $\{r_{\alpha}\}_{\alpha=1}^{N^2}$ with the attached
weights $\{\omega_{\alpha}\}_{\alpha=1}^{N^2}$, then the
estimator (\ref{Ripley}) can be rewritten as
\begin{equation}
\hat{K}(r_{\alpha} \le r < r_{\alpha+1})=\frac{V}{N^2}
\sum_{\beta=1}^{\alpha} \frac{1}{\omega(r_\beta)}.
\label{RO}
\end{equation}
for even indices $\alpha$, since each $r_{\alpha}$ value is
repeated twice.
This is the sample estimate of the
cumulative probability distribution function of pairwise
separations.
For a large point sample the function (\ref{RO}) demands
a huge memory storage. In practice we can approximate this
function by logarithmic discretization of
the total pairwise distance interval.

\begin{figure*}
\centering
\resizebox{.37\textwidth}{!}{\includegraphics*{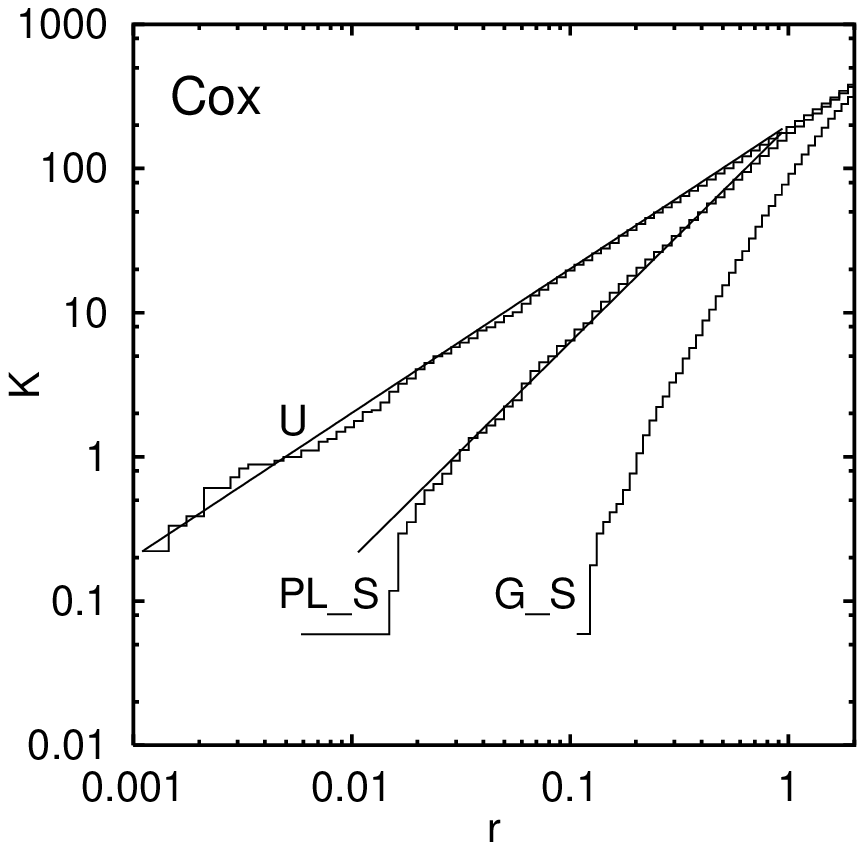}}\quad\quad
\resizebox{.37\textwidth}{!}{\includegraphics*{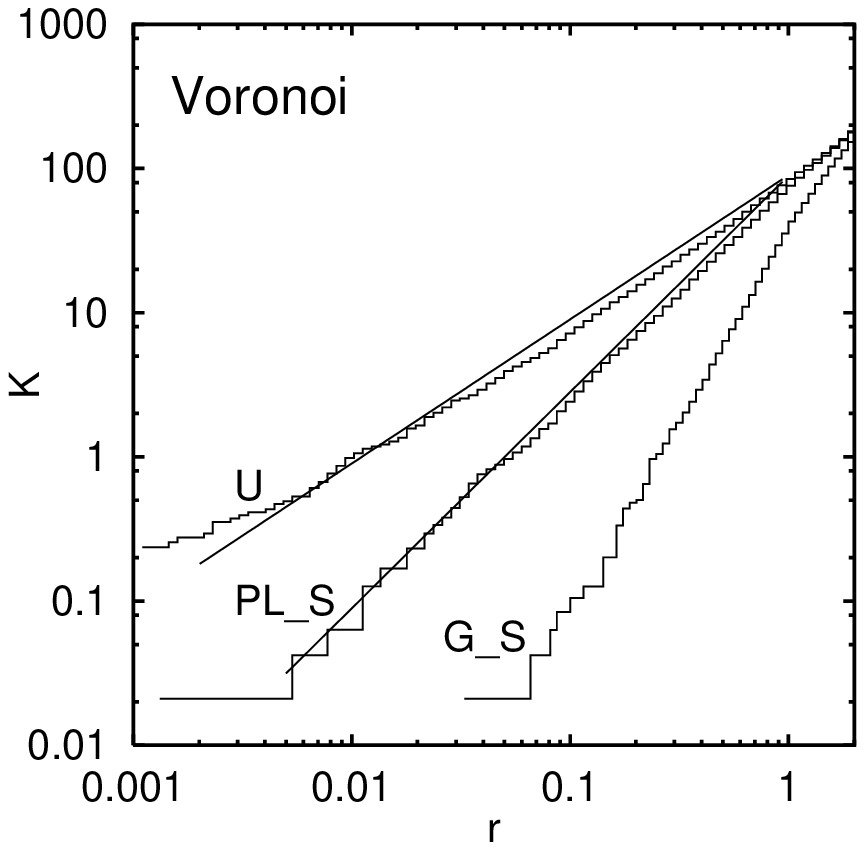}}\\
\caption{Effect of shifting on the correlation integral $K(r)$.
Label U stands for the unshifted original processes, label PL\_S stands
for the power-law
shifted processes, and label G\_S stands for the Gaussian shifted
processes. The left
panel shows the Cox family of processes, the right panel shows the Voronoi
family. The straight
lines represent the expected power-law behaviour.}
\label{kfig}
\end{figure*}

The results are shown in Fig.~\ref{kfig}. The $K$-function is
shown as a staircase, and the straight lines correspond to the
expected correlation dimensions ($D_2=1$ for pure samples,
$D_2=3-\gamma+2(1+\alpha)=1.5$ for shifted samples). We can clearly
see that $D_2$ is extremely close to the expected value. The small
differences between the expected power law and the real $K(r)$
dependence are probably due to the finiteness of the
simulated samples and the algorithmic (precision) problems in
modelling singular point distributions. The correlation integral
for the Gaussian-shifted samples does not reach near the pole and
has the (Poisson) correlation dimension $D_2\approx 3$, indicating
that the correlation structure has been destroyed.

\begin{figure*}
\centering
\resizebox{.39\textwidth}{!}{\includegraphics*{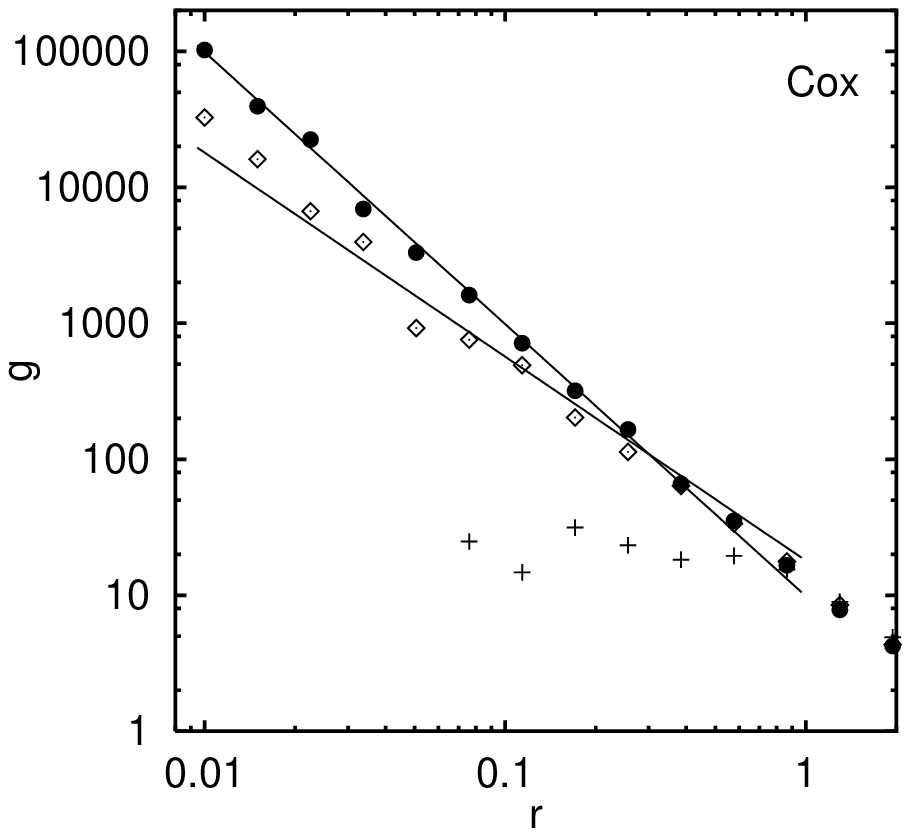}}\quad
\resizebox{.39\textwidth}{!}{\includegraphics*{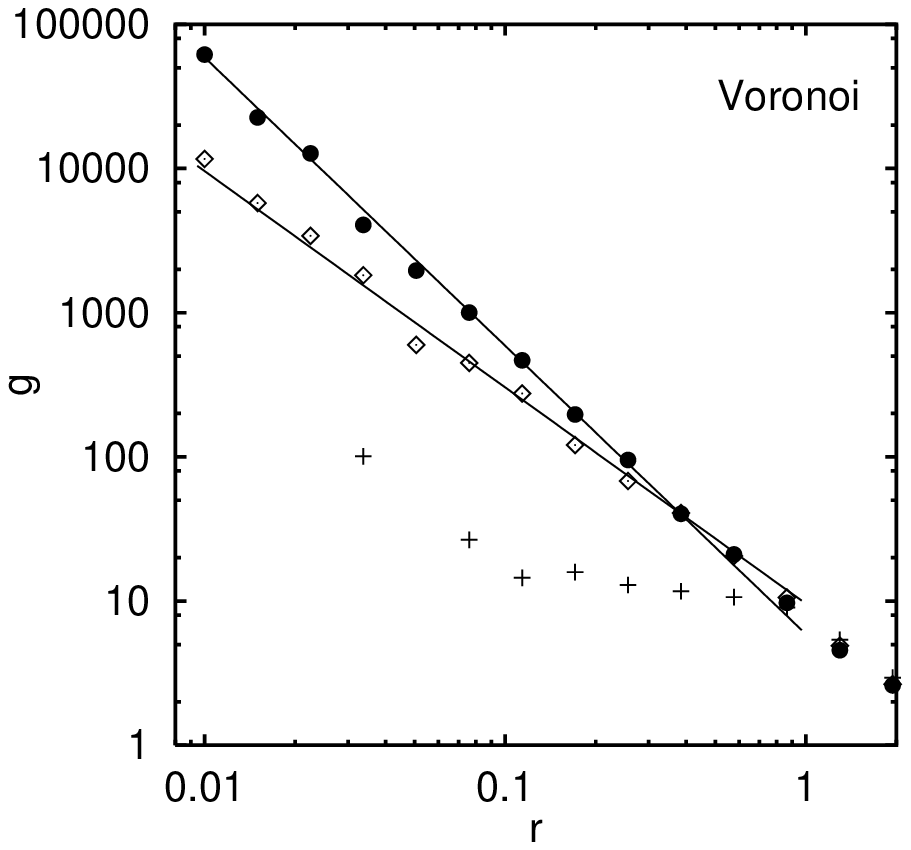}}\\
\caption{Effect of shifting on the correlation function $g(r)$. The left
panel shows the Cox family of processes,
the right panel shows the
Voronoi family. The correlation functions for the original processes are shown
by solid bullets, the power-law shifted processes by open diamonds, and
for Gaussian-shifted processes by crosses.
The straight
lines represent the expected power-law behaviour.}
\label{sose}
\end{figure*}

We have also calculated directly the pair correlation
functions: these do not reach to the smallest pair distances,
as explained above.
Fig.~\ref{sose} shows the pair correlation functions $g(r)=\xi(r)+1$ for the
Cox process and the Voronoi vertices process, both
before and after the random shifts.
We have tried different estimators \citep[see][]{Pons99}, but
no significant differences were found amongst them.
The correlation functions given in Fig.~\ref{sose} were calculated
using the Rivolo estimator \citep{Riv86}.
As in the previous figure,
the straight lines correspond to the expected
order of singularity $\gamma=1.5$.
We can see that $\gamma$ lies in the expected range, supporting
the conjecture that the reduction of the order of the
singularity by $2(1+\alpha)$, using random shifts
with the power-law density $d^*(r)\sim d^\alpha$
is a universal phenomenon.

We also see that Gaussian shifts
have completely destroyed the singularity.

\section{Summary and conclusions}
\begin{enumerate}
        \item
This paper is based on the empirical evidence that the two-point
correlation function $\xi(r)$ of the galaxy point field follows
for small $r$ a power-law $\xi(r)\propto r^{-\gamma}$ with
$1.5<\gamma<1.9$. Nevertheless, physically motivated
published models for the galaxy point pattern do not have this
property.
        \item
A reason for this dilemma may be that these models are
usually based on some simplifying assumptions. As these ideal
conditions are only an approximation of reality, it is not
surprising that also the point field model is an approximation.
 \item
One way to cover the difference between theory and reality is to
add a further random element to the theoretical point field model,
i.\,e. to randomly shift every point.
\item
The determination of the effect of randomly shifting on the order
of the singularity of $\xi(r)$ is too complicated for the general
case. Therefore, this paper has investigated this question only
for a relatively simple class of Poisson cluster processes, the
Neyman--Scott processes. In this case, there is an easy rule how to
choose the distribution of the random shifts to get any desired
reduction of the order of the singularity. If $d^*(r)$, the
distribution of the length of the isotropic shifts, behaves for
small $r$ as $r^\alpha$, then the order is reduced by
$2(1+\alpha)$. Unfortunately, Neyman--Scott processes are not
variable enough to model the galaxy point pattern.
\item
The vertices process of a Poisson--Voronoi tessellation
and the Cox segment process represent the galaxy clustering
better, but their two-point
correlation functions have singularities of order $\gamma=2$. This
is too large. Simulation studies support the
conjecture that randomly shifting
the vertices by power-law distributed random shifts
leads to a reduction of $\gamma$ in the same way as
it does in the case of Neyman--Scott processes.
\item
Thus, a suitably shifted vertices process of a Poisson--Voronoi
tessellation or of the Cox segment process
may be an appropriate model for the galaxy point
pattern, as it is physically convincing and agrees with
observations.
\item
The method of randomly shifting may be applicable also in case of
other initial point fields, apart from the processes
studied here. This may
lead to a number of reasonable models for the galaxy point pattern.
\end{enumerate}
\subsection*{Acknowledgments}
We thank the referee, Stephane Colombi, for his constructive
criticism and useful suggestions.
This work was supported by the Spanish MCyT project AYA2000-2045.
ES acknowledges support from the Vicerrectorado de Investigaci\'on
de la Universitat de Val\`encia.

\appendix
\section{Derivation of Eq.\ (\ref{inti})}
Here, the asymptotic behaviour of $f^*(r)$ for $r\to0$ is
determined. The starting point is the integral (\ref{inti})
\begin{equation}
        F^*(r)=\int\limits_{-\infty}^\infty\int\limits_{-\infty}^\infty f(x)d^*(z)
        \frac{\textrm{A}(x;z;r)}{4\pi z^2}\textrm{d}x\textrm{d}z,
\end{equation}
where $f(r)$ behaves as $r^{2-\gamma}$ (i.\,e. the order of the
singularity of $g(r)$ before the random shifts is $\gamma$) and $d^*(r)$
behaves as $r^\alpha$, $\gamma<3$ and $\alpha>-1$.

Elementary considerations show that the asymptotic behaviour of
$F^*(r)$ is equivalent to the case where
$f(r)=r^{2-\gamma}\Beins_{\lbrack0,1\rbrack}(r)$ and
$d^*(r)=r^\alpha\Beins_{\lbrack0,1\rbrack}(r)$, 
where $\Beins_{\lbrack a,b\rbrack}(r)$ is the indicator function
of the interval $\lbrack a,b\rbrack$,
\begin{equation}
\Beins_{\lbrack a,b\rbrack}(r) =\left\{
        \begin{array}{r@{\quad}l}
        1 & \mbox{for } r \in \lbrack a,b\rbrack ,\\
        0 & \mbox{for } r \notin  \lbrack a,b\rbrack.
        \end{array}
        \right.
\end{equation}
Thus, the calculations are carried out only for this case.

We can write
\begin{eqnarray}
        F^*(r) & = & \hspace{0.5cm}\int\limits_{0}^{r}\int\limits_{r-z}^{r+z}
        \frac{r^2-(x-z)^2}{4x^{\gamma-1}z^{1-\alpha}}\textrm{d}x\textrm{d}z
        \nonumber \\
        & & {}+\int\limits_{r}^{1}\int\limits_{z-r}^{z+r}
        \frac{r^2-(x-z)^2}{4x^{\gamma-1}z^{1-\alpha}}\textrm{d}x\textrm{d}z
        \nonumber \\
        & & {}+\int\limits_{0}^{r}\int\limits_{0}^{r-z}
        x^{2-\gamma} z^\alpha\textrm{d}x\textrm{d}z,
\end{eqnarray}
and its derivative
\begin{eqnarray}
        f^*(r) & = & \hspace{0.5cm}
        \int\limits_{0}^{r}\int\limits_{r-z}^{r+z}
        \frac{r}{2x^{{\gamma-1}}z^{1-\alpha}}\textrm{d}x\textrm{d}z+
        \frac{r^{3-\gamma+\alpha}}{(3-\gamma)(4-\gamma)} \nonumber \\
        & & {}-\frac{\Gamma(1+\alpha)\Gamma(3-\gamma)}
{\Gamma(4+\alpha-\gamma)}r^{3-\gamma+\alpha} \nonumber \\
        & & {}+\int\limits_{r}^{1}\int\limits_{z-r}^{z+r}
        \frac{r}{2x^{\gamma-1}z^{1-\alpha}}\textrm{d}x\textrm{d}z
 -\frac{r^{3-\gamma+\alpha}}{2^{\gamma-2}(3-\gamma)} \nonumber\\
        & & {}+\frac{\Gamma(1+\alpha)\Gamma(3-\gamma)}
        {\Gamma(4+\alpha-\gamma)}r^{3-\gamma+\alpha} \nonumber \\
        & = & \hspace{0.5cm}\frac{r}{2(2-\gamma)}\int\limits_{0}^{r}
        \frac{(r+z)^{2-\gamma}-(r-z)^{2-\gamma}}{z^{1-\alpha}}\textrm{d}z 
        \nonumber \\
        & & {}+\frac{r}{2(2-\gamma)}\int\limits_{r}^{1}
        \frac{(z+r)^{2-\gamma}-(z-r)^{2-\gamma}}{z^{1-\alpha}}\textrm{d}z 
        \nonumber \\
        & & {}+\frac{r^{3-\gamma+\alpha}}{3-\gamma}\left(
        \frac{1}{4-\gamma}-2^{2-\gamma}\right) \nonumber \\
        & = & \hspace{0.5cm}\frac{r}{2(2-\gamma)}
        \int\limits_{0}^{1}\frac{(r+z)^{2-\gamma}}{z^{1-\alpha}}\textrm{d}z 
        \nonumber \\
        & & {}-\frac{r}{2(2-\gamma)}\int\limits_{0}^{r}
        \frac{(r-z)^{2-\gamma}}{z^{1-\alpha}}\textrm{d}z \nonumber \\
        & & {}-\frac{r}{2(2-\gamma)}\int\limits_{r}^{1}
        \frac{(z-r)^{2-\gamma}}{z^{1-\alpha}}\textrm{d}z \nonumber \\
        & & {}+\frac{r^{3-\gamma+\alpha}}{3-\gamma}\left(
        \frac{1}{4-\gamma}-2^{2-\gamma}\right) \nonumber \\
        & = & \hspace{0.5cm}\frac{r}{2(2-\gamma)}
        {}_2\mathrm{F}_1(\gamma-2,\alpha,1+\alpha,-1/r)r^{2-\gamma}/\alpha 
        \nonumber \\
        & & {}-\frac{r}{2(2-\gamma)}\frac{\Gamma(3-\gamma)\Gamma(\alpha)}
        {\Gamma(3-\gamma+\alpha)}r^{2-\gamma+\alpha} \nonumber \\
        & & {}-\frac{r}{2(2-\gamma)}\frac{(1-r)^{3-\gamma}}{3-\gamma} 
        \nonumber \\
        & & {}\times{}_2\mathrm{F}_1(3-\gamma,1-\alpha,4-\gamma,1-1/r)r^{-1+\alpha} 
        \nonumber \\
        & & {}+\frac{r^{3-\gamma+\alpha}}{3-\gamma}\left(
        \frac{1}{4-\gamma}-2^{2-\gamma}\right),
\end{eqnarray}
where ${}_2\mathrm{F}_1$ is the Gaussian hypergeometrical
function \citep[see][]{Kratzer60}.

For small $r$ it behaves as
\begin{eqnarray}
        \lefteqn{{}_2\mathrm{F}_1(\gamma-2,\alpha,1+\alpha,-1/r)} \nonumber \\
 & \sim &
        \left\{\begin{array}{lll}
                \frac{\displaystyle\Gamma(1+\alpha)\Gamma(\gamma-2-\alpha)}
                {\displaystyle\Gamma(\gamma-2)}r^\alpha,
                & \mathrm{if} & 2-\gamma+\alpha<0,\\
                \frac{\displaystyle\alpha}{\displaystyle2-\gamma+\alpha}
                r^{\gamma-2}, & \mathrm{if} &
                2-\gamma+\alpha\geq0,
        \end{array}\right.
\end{eqnarray}
and
\begin{eqnarray}
        \lefteqn{{}_2\mathrm{F}_1(3-\gamma,1-\alpha,4-\gamma,1-1/r)} \nonumber \\
 & \sim &
        \left\{\begin{array}{lll}
                \frac{\displaystyle\Gamma(4-\gamma)\Gamma(\gamma-2-\alpha)}
                {\displaystyle\Gamma(1-\alpha)(1-r)^{3-\gamma}}r^{3-\gamma},
                & \mathrm{if} & {2-\gamma}+\alpha<0,\\
                \frac{\displaystyle{3-\gamma}}
                {\displaystyle({2-\gamma}+\alpha)(1-r)^{1-\alpha}}
                r^{1-\alpha}, & \mathrm{if} & {2-\gamma}+\alpha\geq0.
        \end{array}\right.
\end{eqnarray}
When ${2-\gamma}+\alpha<0$, all terms behave as
$r^{3-\gamma+\alpha}$. When ${2-\gamma}+\alpha\geq0$,
all linear terms cancel each other out. Thus, in any case there remain only
terms which behave as $r^{3-\gamma+\alpha}$
and so does $f^*(r)$.

\end{document}